\documentclass[a4paper,twocolumn,10pt]{quantumarticle}
\pdfoutput=1
\usepackage[T1]{fontenc}
\usepackage[numbers,sort&compress]{natbib}
\usepackage{color}
\usepackage{bbold}
\usepackage{dsfont}
\usepackage{graphicx}
\usepackage[caption=false]{subfig}
\usepackage[colorlinks]{hyperref}
\usepackage[bottom]{footmisc}
\usepackage{enumerate}
\usepackage{float}
\usepackage{dblfloatfix}
\usepackage{mathtools}

\usepackage{empheq}
\usepackage{tikz}
\usepackage{fancyhdr}
\usepackage[normalem]{ulem}
\DeclareMathAlphabet\mbc{OMS}{cmsy}{b}{n}
\usepackage{balance}

\usepackage{multirow}
\usepackage{colortbl}
\usepackage{lipsum}
\usepackage[most]{tcolorbox}

\begin{document}

\global\long\def\eqn#1{\begin{align}#1\end{align}}
\global\long\def\vec#1{\overrightarrow{#1}}
\global\long\def\ket#1{\left|#1\right\rangle }
\global\long\def\bra#1{\left\langle #1\right|}
\global\long\def\bkt#1{\left(#1\right)}
\global\long\def\sbkt#1{\left[#1\right]}
\global\long\def\cbkt#1{\left\{#1\right\}}
\global\long\def\abs#1{\left\vert#1\right\vert}
\global\long\def\cev#1{\overleftarrow{#1}}
\global\long\def\der#1#2{\frac{{d}#1}{{d}#2}}
\global\long\def\pard#1#2{\frac{{\partial}#1}{{\partial}#2}}
\global\long\def\re{\mathrm{Re}}
\global\long\def\im{\mathrm{Im}}
\global\long\def\dd{\mathrm{d}}
\global\long\def\ddd{\mathcal{D}}

\global\long\def\avg#1{\left\langle #1 \right\rangle}
\global\long\def\mr#1{\mathrm{#1}}
\global\long\def\mb#1{{\mathbf #1}}
\global\long\def\mc#1{\mathcal{#1}}
\global\long\def\tr{\mathrm{Tr}}
\global\long\def\dbar#1{\Bar{\Bar{#1}}}

\global\long\def\nth{$n^{\mathrm{th}}$\,}
\global\long\def\mth{$m^{\mathrm{th}}$\,}
\global\long\def\non{\nonumber}

\newcommand{\orange}[1]{{\color{orange} {#1}}}
\newcommand{\teal}[1]{{\color{teal} {#1}}}
\newcommand{\cyan}[1]{{\color{cyan} {#1}}}
\newcommand{\blue}[1]{{\color{blue} {#1}}}
\newcommand{\yellow}[1]{{\color{yellow} {#1}}}
\newcommand{\green}[1]{{\color{green} {#1}}}
\newcommand{\red}[1]{{\color{red} {#1}}}
\newcommand{\pbb}[1]{{\textcolor{teal}{[PBB: #1]}}}

\global\long\def\todo#1{\orange{{$\bigstar$ \cyan{\bf\sc #1}}$\bigstar$} }
\newcommand{\ks}[1]{{\textcolor{teal}{[KS: #1]}}}

\title{Dissimilar collective decay and directional emission from two quantum emitters}

\author{P. Solano}
\email{psolano@udec.cl}
\affiliation{Departamento de F\'isica, Facultad de Ciencias F\'isicas y Matem\'aticas, Universidad de Concepci\'on, Concepci\'on, Chile}

\author{P. Barberis-Blostein}
\affiliation{Instituto de Investigaciones en Matem\'{a}ticas Aplicadas y en Sistemas, Universidad Nacional Aut\'{o}noma de M\'{e}xico, Ciudad Universitaria, 04510, DF, M\'{e}xico.}

\author{K. Sinha}
\email{kanu.sinha@asu.edu}
\affiliation{School of Electrical, Computer and Energy Engineering, Arizona State University, Tempe, AZ 85287-5706, USA}

\begin{abstract}
We study a system of two distant quantum emitters coupled via a one-dimensional waveguide where the electromagnetic field has a direction-dependent  velocity. As a consequence, the onset of collective emission is non-simultaneous and, for appropriate parameters, while one of the emitters exhibits superradiance the other can be subradiant. Interference effects enable the system to radiate in a preferential direction depending on the atomic state and the field propagation phases. We characterize such directional emission as a function of various parameters, delineating the  conditions for optimal directionality.

\end{abstract}

\maketitle

\section{Introduction}
Engineering atom-photon interactions by manipulating electromagnetic (EM) fields is a significant aspect of design and implementation of quantum technologies
\cite{PierreBook2, MitchellBook, QTMilburn}.  For instance, reducing the mode volume of the EM field enhances the light-matter coupling \cite{HarocheBook} and controlling the field polarization allows for chiral interactions between quantum emitters with polarization-dependent transitions \cite{Lodahl17}. Current platforms allow one to change yet another property of the EM field: its propagation velocity \cite{Hood2016,Liu2017,Pedersen2008}. In particular, one can envision the possibility of having an EM field with unequal velocities when propagating to the left or to the right, here referred to as \textit{anisotachy} \footnote{The term `\textit{anisotachy}' comprises of the Greek words `\textit{aniso}' for unequal and `\textit{tachytita}' for velocity.}. Such feature is, as yet, an unexplored aspect of quantum optical systems, which could be implemented with state-of-the art non-reciprocal components \cite{Asadchy2020,Jala2013}. Since the propagation velocity is an essential ingredient in connecting the distant parts of a larger system, the effects of anisotachy are expected to appear when measuring properties that depend on the interaction between delocalized subsystems, such as quantum correlations.

Quantum correlations among   emitters can collectively enhance or inhibit light absorption and emission \cite{Gross82}. For example, a collection of emitters can radiate faster or slower than individual ones depending on their correlations, phenomena known as super- and sub-radiance respectively \footnote{In the context of this paper,  super- and sub-radiance indicate the instantaneous rate of atomic excitation decay being faster or slower than independent decay.}. These effects have been extensively studied both theoretically \cite{Dicke,Gross82,Rehler71,Eberly72,Asenjo17,Needham19} and experimentally across various platforms \cite{Skribanowitz73,Pavolini85,Gross76,Devoe96,Scheibner07,Rohlsberger10,Mlynek14,Goban15,Guerin16,Bradac17,Solano2017,Wang20,Ferioli21}. Recent works have proposed, and implemented \cite{Kannan2022}, collective effects for controlling the direction of emission using the non-local correlations between two emitters, with potential applications in quantum information processing and quantum error correction \cite{Guimond20,Gheeraert20,Yang21,Du21}

\begin{figure}[b]
    \centering
    \includegraphics[width = 3.2 in]{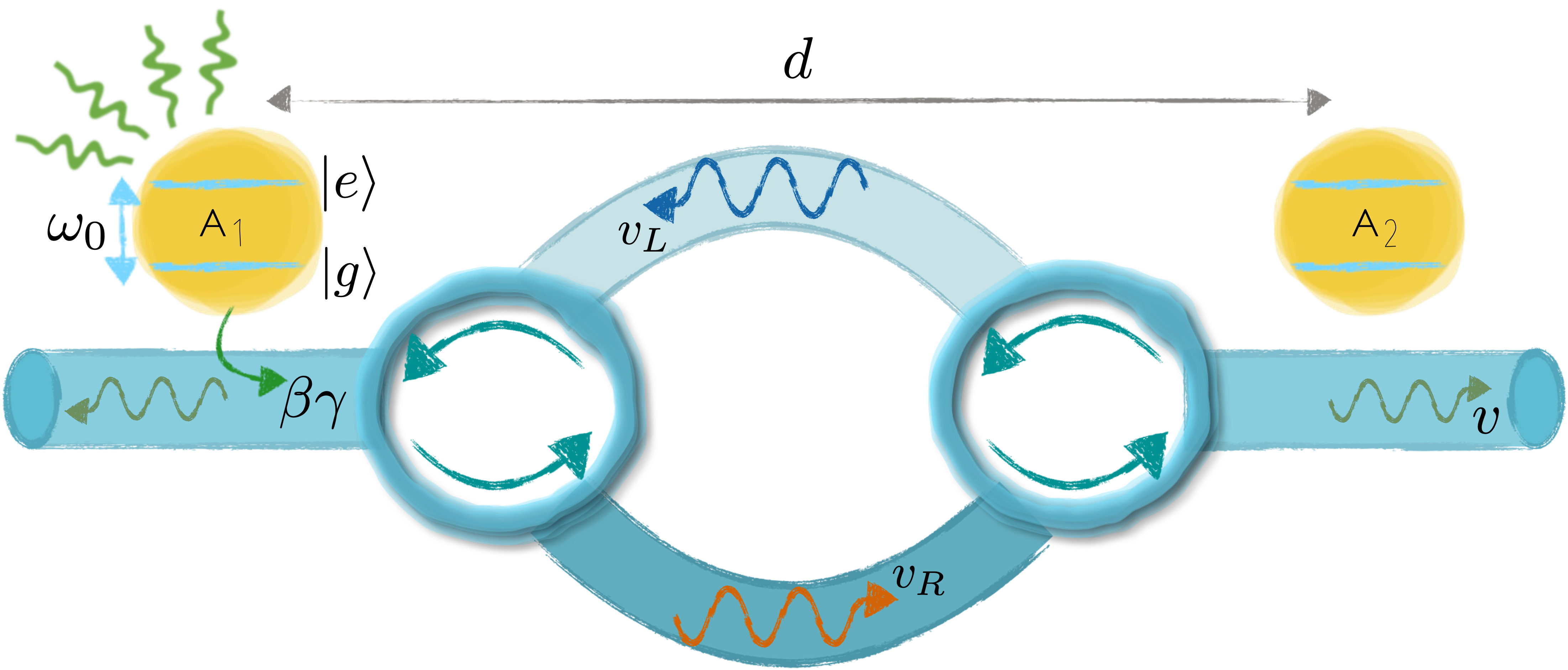}
    \caption{Schematic representation of two two-level atoms coupled to a waveguide with an interatomic separation $d$. The waveguide connects to two circulators that are coupled to two separate waveguides with different  field velocities, such that the left (right) propagating modes have a velocity  $ v_{\rm{L}}$ $(v_{\rm{R}})$. The atoms have a resonance frequency $\omega_0$ and a decay rate $\gamma$, and couple to the waveguide with a coupling efficiency $\beta$. The velocity of the field in the waveguide outside of the circulator region is $v$.}
    \label{Fig:sch}
\end{figure}

In this work we propose a system comprising of two distant quantum emitters or atoms coupled to a one-dimensional waveguide with an effective direction-dependent field velocity, or anisotachy. As we will show, a direction-dependent time delay can allow  two correlated emitters to exhibit disparate collective effects such that while one atom decays superradiantly, the other exhibits subradiance. In such a system, interference effects in the radiated field lead to a directional emission. We characterize such directional emission as a function of initial states of the emitters, field propagation phases, and the waveguide coupling efficiency.  Our results demonstrate that collective directional emission is a rather prevalent quantum optical phenomenon that needs further exploration, to understand its advantages, limitations and dependence on a broader set of parameters. 

We first present a theoretical model of the system, describing the disparate cooperative decay dynamics of the emitters  and the radiated field intensity in the presence of anisotachy. We then characterize the optimum conditions for directional emission. Finally we discuss the experimental feasibility and give a brief outlook of the phenomena.

\begin{figure}[t]
\centering
\subfloat[]{\includegraphics[width=2.7 in]{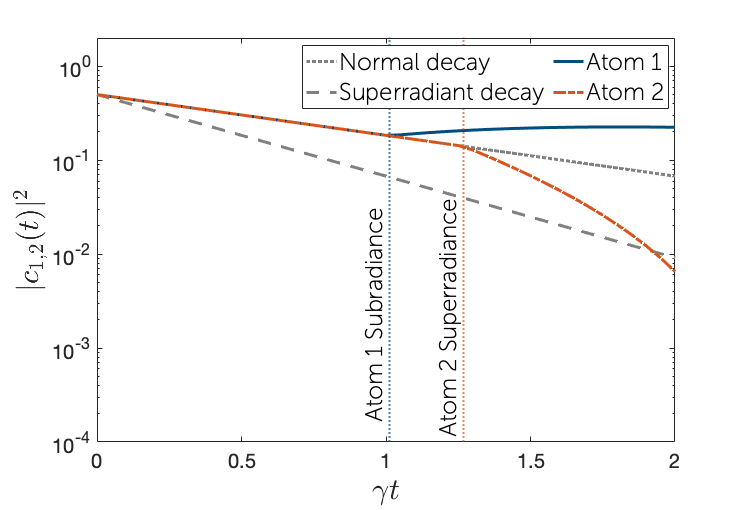}}\\
\subfloat[]{\includegraphics[width=2.8 in]{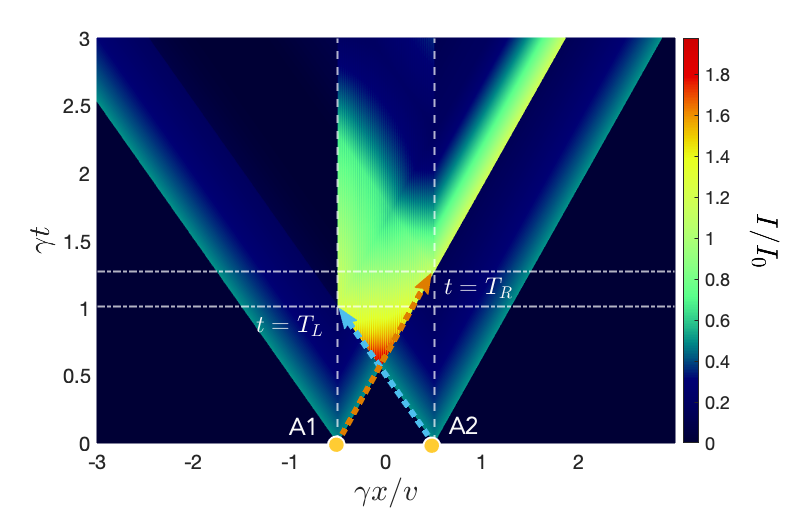}}
\caption{Atomic excitation probability and intensity dynamics for emitters prepared initially in the symmetric state $\ket{\Phi_0 }= \frac{1}{\sqrt 2}\bkt{ \ket{eg} + \ket{ge}}$, with   $\phi_{\rm{L}} = \pi$ and  $\phi_{\rm{R}} = 2\pi$. (a) The blue solid and red dash-dotted curves represent the populations of atoms A1 and A2 respectively, with the dotted vertical lines indicating the onset of collective emission.  The gray dotted and dashed curves represent standard  individual and  superradiant decay, respectively. (b) Intensity of the radiated field as a function of spacetime. The vertical dashed lines indicate the position of the two atoms while the horizontal dash-dotted lines represent the times $t = T_L$ and $t = T_R$. Here we have chosen the dimensionless atomic separation $\gamma d/v\approx1$,  propagation velocities  $ v_{\rm{L}}/v \approx 0.988$ and  $v_{\rm{R}} /v\approx 0.788$,  $\omega_0/\gamma \approx 500$ and coupling efficiency $\beta \approx1$. }
\label{fig:decays}
\end{figure}
\section{Model}
\label{Sec: Model}
Let us consider two two-level quantum emitters coupled to the EM field modes of a waveguide. Using field circulators, the field modes propagate through different waveguides with unequal index of refraction, as shown in Fig. \ref{Fig:sch}, leading to an effective anisotachy.

The Hamiltonian for the system is given as $
H = H_E  + H_F + H_{EF}$, where $H_E  = \sum_{m = 1, 2} \hbar  \omega_0 \hat \sigma_m^+ \hat \sigma_m^ -$ is the Hamiltonian for the emitters, with $\cbkt{\hat \sigma_m^+, \hat \sigma_m^-} $ as the raising and lowering operators for the $m^{\mr{th}}$ atom. $
H_F = \int_0 ^\infty \dd\omega \,\hbar \omega \sbkt{\hat{a}_{\mr{R}}^\dagger\bkt{\omega} \hat{a}_{\mr{R}}\bkt{\omega} + \hat{a}_{\mr{L}}^\dagger\bkt{\omega} \hat{a}_{\mr{L}}\bkt{\omega}},$ corresponds to the Hamiltonian for the guided modes of the waveguide, with $\hat a_{\rm{L(R)}}^{(\dagger)} $ as the bosonic operators for the left (right) propagating modes. The interaction Hamiltonian in the interaction picture with respect to the free Hamiltonians $H_E + H_F$ is \cite{PeterQOBook, PierreBook}
\eqn{\tilde H_{EF} =& \sum_{m = 1,2} \int_0 ^\infty \dd \omega\,\hbar g\bkt{\omega}\sbkt{\hat{\sigma}_m^+\cbkt{\hat {a}_{\rm{R}}\bkt{\omega}e^{i k_{\rm{R}} x_m } \right.\right.\non\\
&\left.\left.+ \hat {a}_{\rm{L}}\bkt{\omega}e^{-i k_{\rm{L}} x_m }}e^{ - i \bkt{\omega - \omega_0 }t} + \rm{H.C.}},
\label{Eq:hef}
}
where $x_1 = -x_2 = -d/2 $ denotes the position of the emitters, $g(\omega) $ represents the atom-field coupling strength and $k_{L, R}\equiv \omega/v_{L,R}$ corresponds to the asymmetric left and right wavenumbers. Considering the initial state of the system with the emitters being in the single excitation sub-space and the field in vacuum,
\eqn{\label{Eq:psit}
\ket{\Psi(t)} = \sbkt{\sum_{m = 1,2} c_m (t) \hat{\sigma}_m^+ + \int_0 ^\infty \dd\omega \, \cbkt{c_{\rm{R}} (\omega,t)  \hat{a}_{\rm{R}}^\dagger\bkt{\omega}\right.\right. &\non\\
\left.\left.+ c_{\rm{L}} (\omega,t)  \hat{a}_{\rm{L}}^\dagger\bkt{\omega}}}\ket{gg}\otimes \ket{\cbkt{0 }}&,
}
one can derive the equations of motion for the atomic coefficients, $c_1$ and $c_2$,  using a Wigner-Weisskopf approach as  (see Appendix\,\ref{Appen A} for details)
\eqn{
\dot{c}_{1(2)}\bkt{t}  = -& \frac{\gamma}{2} \sbkt{c_{1(2)}\bkt{t}\right.\non\\
&\left.+\beta c_{2(1)} \bkt{t-T_{\rm{L(R)}}}\Theta\bkt{t-T_{\rm{L (R)}}}e^{i \omega_0 T_{\rm{L(R)}}}} \label{diffeq:12}.
}
Here $\gamma$ is the total spontaneous emission rate, $\beta \gamma=4\pi |g(\omega_0)|^2$ is the decay rate into the guided modes, and $T_{\rm{R}(\rm{L})}=d/v_{\rm{R}(\rm{L})}$ is the propagation time of the field traveling right (left) from one emitter to the other.

\section{Dynamics}
\label{Sec: Dynamics}
Let us consider  the initial state of the emitters to be $\ket{\Psi_0} \equiv c_1(0) \ket{eg}+c_2(0) \ket{ge}$. The  equations of motion  (Eq.\,\eqref{diffeq:12}) can be solved to obtain (see Appendix\,\ref{Appen. B}): 
\begin{widetext} 
\eqn{ c_{1(2)} (t) = &\sum_{n = 0}^\infty\underbrace{\boxed{c_{1 (2)}(0)e^{i2n \phi}}\frac{\bkt{\beta \gamma /2}^{2n}}{(2n)!}   \bkt{t - 2n T }^{2n} e^{- \gamma  \bkt{ t - 2nT}/2} \Theta\bkt{t - 2 n T}}_{\text{Even number of trips between atoms}} \non\\
& -\sum_{n = 0}^\infty\underbrace{\boxed{c_{2 (1)}(0)e^{i2n\phi + i \phi_{L(R)}}}\frac{\bkt{\beta \gamma /2}^{2n+1}}{(2n+1)!}  \bkt{t - 2n T - T_{L(R)} }^{2n+1} e^{- \gamma  \bkt{ t - 2nT- T_{L(R)}}/2} \Theta\bkt{t - 2 n T- T_{L(R)}}}_{\text{Odd number of  trips between atoms}},\label{eq:c12}
}
\end{widetext}
where $\phi_{\rm{R}(\rm{L})}=\omega_0 T_{\rm{R}(\rm{L})}$ is the phase acquired by the resonant field upon propagation between the emitters, and $T=(T_{\rm{R}}+T_{\rm{L}})/2$ and $\phi=(\phi_{\rm{R}}+\phi_{\rm{L}})/2$ are the average propagation time and phase respectively \footnote{In the absence of anisotachy, it simplifies to previous results of collective radiation in the presence of delay \cite{superduper1,SPIE,Sinha20, Dinc18, Dinc19}}.  The first term in the equation above represents the modification of atomic decay after $n$ round trips of the field between the emitters. The second term represents an odd-number of trips $(2n+1)$ from one emitter to the other. The directional propagation phase ($\phi_{\rm{R}(\rm{L})}$), together with the phase from the atomic coefficients ($c_{2(1)}$), determines the constructive or destructive nature of interference between the two terms, as indicated by the boxed terms.

Fig.\,\ref{fig:decays} (a) shows the decay of the atomic excitation coefficients as a function of time. Destructive (constructive) interference in the left (right) propagating modes leads to a subradiant atom A1 and a superradiant atom A2, after the field from one atom reaches the other. One can thus interpret collective decay as a mutually stimulated emission process, as is evident from the series expansion in Eq.\,\eqref{eq:c12}.  For a negligible separation between emitters $(T\rightarrow0)$, the series converges to yield the standard superradiant exponential decay. In the presence of delay, the resulting dynamics is more precisely described as a cascade of stimulated emission processes \cite{Milonni74}. For instance, the field from one emitter can stimulate emission of the other, leading to a non-exponential decay that is faster than superradiance \cite{superduper1, Dinc18,Dinc19,Longhi2020}, or completely suppress its emission, leading to bound states in the continuum (BIC) \cite{SPIE,Calajo19}. More generally, this effect can accelerate the decay of one atom while slowing the decay of the other, as shown in Fig.\,\ref{fig:decays} (a). This demonstrates that the phenomena of super- and subradiance are not a characteristic of the system as a whole, rather an effective description of local atom-photon interference effects.
\begin{widetext}
\begin{center}

\begin{table}[t]

\centering
\begin{tabular}{ |c|c|c|c|c|c| } 
\hline
$\phi_L$& $\phi_R$ & $\phi_{A1} - \phi_{A2}$  & Atom 1& Atom 2& \\
\hline
\multirow{2}{5em}{$2n\pi$}& \multirow{2}{5em}{$(2m+1)\pi$} & $0 $ &\cellcolor{yellow!50} Superradiant & \cellcolor{gray!50}Subradiant& Anisotachy\\ 
& & $\pi$ &\cellcolor{gray!50}Subradiant &\cellcolor{yellow!50}Superradiant&required \\ 
\hline
\multirow{2}{5em}{$(2n+1)\pi$}& \multirow{2}{5em}{$2m\pi$} & $0 $&\cellcolor{gray!50}Subradiant &\cellcolor{yellow!50}Superradiant & Anisotachy  \\ 
& & $\pi$  & \cellcolor{yellow!50}Superradiant & \cellcolor{gray!50}Subradiant&required\\ 
\hline
\multirow{2}{5em}{$(n+\frac{1}{2})\pi$}& \multirow{2}{5em}{$(m+\frac{1}{2})\pi$} & $\frac{\pi}{2} $& \cellcolor{gray!50}Subradiant &\cellcolor{yellow!50}Superradiant& No anisotachy   \\ 
& & -$\frac{\pi}{2}$  & \cellcolor{yellow!50}Superradiant & \cellcolor{gray!50}Subradiant&required \\ 
\hline 
\end{tabular}
\caption{Some representative examples of dissimilar collective decay for different atomic and field propagation phases.}
\label{table_phases}
\end{table}
\end{center}
\end{widetext}

One can note a few salient features of the collective atomic dynamics from the above equation:
\begin{itemize}
    \item {Each term in the series expansion can be interpreted as multiple partial reflections of a field wave packet bouncing between the atoms at signaling times $t = 2nd/v$ and $2nd/v +d/v_{L(R)}$, as denoted by the theta-functions ($\Theta \bkt{t - 2nd/v}$ and $\Theta \bkt{t - 2nd/v-d/v_{L(R)}}$). This offers the intuition that the collective decay dynamics arises from a cascade of stimulated emission processes as the field emitted by the each of the atoms propagates back and forth between them.}

    \item{The interference phase for all  partial reflections is determined by the phase factors $c_{1(2)}(0) e^{i 2n \phi}$ and $c_{2(1)}(0) e^{i 2n \phi + i \phi_{L(R)}}$,  which is a combination of the atomic and field propagation phases. Each successive term comes with an additional factor of the atom-waveguide coupling strength $\beta\gamma/2$.}
    
    \item{The propagation phases $\phi_{L(R)}$ can be different in the presence of anisotachy, which can make the contribution from the second term to the collective dynamics different for the two atoms, thus leading to  dissimilar collective behavior. We summarize a few example cases of such behavior in Table~\ref{table_phases}}.
\end{itemize}

The EM field intensity emitted by the system, as a function of
position $x$ and time $t$, can be evaluated as $I\bkt{x,t} =
\frac{\epsilon_0 c} {2}\avg{\Psi\bkt{t}\vert\hat{E}^\dagger\bkt{x,t}
  \hat E\bkt{x,t}\vert\Psi\bkt{t}}$ \footnote{Here $\hat{E}\bkt{x,t} =
\int_0^\infty  \dd k\,\mc{E}_k \sbkt{\hat{a}_{\rm{L}} \bkt{k}e^{-i
    k_{\rm{L}} x} + \hat{a}_{\rm{R}} \bkt{k}e^{i k_{\rm{R}} x}}e^{-i
  \omega t}$ is the electric field operator, and we have assumed $\mc{E}_k\approx\mc{E}_{k_0}$ to be constant  near the atomic resonance frequency.}  (see Appendix\,\ref{Appen. C}):
\begin{widetext}
\eqn{&\frac{I(x,t)}{ I_0} =\abs{  \sbkt{\underbrace{C_1\bkt{t,\bkt{x +d/2}/v_{\rm{L}} } e^{-i\omega_0 \bkt{x +d/2}/v_{\rm{L}} }}_{\text{Atom 1 left light cone}}+\underbrace{C_2\bkt{ t,\bkt{x - d/2}/v_{\rm{L}}}  e^{-i \omega_0 \bkt{x - d/2}/v_{\rm{L}} }}_{\text{Atom 2 left light cone}}\right.\right.\non\\
 &\left.\left. +\underbrace{C_1\bkt{t,-\bkt{x +d/2}/v_{\rm{R}}} e^{i \omega_0 \bkt{x +d/2}/v_{\rm{R}} }}_{\text{Atom 1 right light cone}} +\underbrace{C_2\bkt{t,-\bkt{x - d/2}/v_{\rm{R}}}  e^{i \omega_0 \bkt{x - d/2}/v_{\rm{R}} }}_{\text{Atom 2 right light cone}}}}^2,
 \label{eq:int}
}
\end{widetext}
where we have redefined the atomic excitation coefficients $C_i\bkt{t ,\tau}= c_i\bkt{t  + \tau}\zeta \bkt{t ,\tau} $  to explicitly include the causal dynamics in the notation, with  $ \zeta (t_1, t_2)\equiv \Theta (t_1 + t_2) - \Theta (t_2)$. The first (last) two terms above correspond to the left-(right-) going light cones emitted from the atoms A1 and A2.  Fig.\,\ref{fig:decays} (b) shows the radiated intensity with the emitted fields destructively (constructively) interfering to the left (right) leading to almost perfect directional emission.

\section{Directional emission}
\label{Sec: Directional emission}
We characterize the probability of emitting the photon into the right
(left) propagating mode by $P_{\rm{R(L)}
}(\Psi_0)=\lim_{t\rightarrow\infty } \int _0 ^\infty\dd\omega
\abs{c_{\rm{R(L)}}\bkt{\omega, t}}^2$, with $P_{\rm{R(L)} }(\Psi_0)$
as an explicit function of the initial state $\ket{\Psi_0} $ (see Appendix\,\ref{Appen. D}).
We focus here on the limit of small atomic separation $\gamma T \ll 1$
such that retardation effects are negligible.  For convenience
  we write the initial condition as $c_1(0)=\cos\theta
  e^{i\phi_{\rm{A1}}}$ and  $c_2(0)=\sin \theta e^{i \phi_{\rm{A2}}}$. The probability of emitting in a particular direction is a function of four parameters: the coupling efficiency $\beta$, the average propagation phase $\phi$; the initial atomic populations parametrized by $\theta$; and the difference between the relative atomic and propagation phases $\Delta \phi= (\phi_{\rm{A1}}-\phi_{\rm{A2}})+(\phi_{\rm{R}}-\phi_{\rm{L}})/2$. For a given experimental realization, the parameters $\beta$ and $\phi$ would be fixed, and the variable atomic parameters $\theta$ and $\Delta\phi$ would determine the directionality of emission.

The directional emission of the system can be characterized by
$\chi=\bkt{P_{\rm{R}}-P_{\rm{L}}}/\bkt{P_{\rm{R}}+P_{\rm{L}}}$, which can be calculated explicitly as (see Appendix\,\ref{Appen. D})
\eqn{
\label{Eq:chi}
\chi(\Psi_0)=-\frac{\beta\sin\phi}{P_{\rm{tot}}}\bkt{\frac{\sin\Delta\phi\sin2\theta+\beta\cos2\theta\sin\phi}{1+\left(\beta\sin\phi\right)^{2}}},
}
where $P_\mr{tot} = P_\mr{R} + P_\mr{L}$ is the total probability of emitting into the waveguide:

\eqn{
\label{eq:ptot}
P_{\rm{tot}}(\Psi_0)&=\non\\
&\beta\sbkt{\frac{1-\beta\cos^{2}\phi-\left(\beta-1\right)\cos\Delta\phi\cos\phi\sin2\theta}{1-\left(\beta\cos\phi\right)^{2}}}.
}
We note from the above that $P_\mr{tot}=\beta$ only for $\phi=\bkt{n+\frac{1}{2}}\pi$, more generally, the interference in the field enhances or inhibits the effective coupling efficiency between the emitters and the waveguide. Eq. \eqref{Eq:chi} shows that the emission is typically directional for most values of $\Delta\phi$ and $\phi\neq N\pi$, $N\in \mathbb{N}$. This indicates a prevalence of directional emission in quantum optical systems.


\begin{figure}[t]
    \centering
    \subfloat[]{\includegraphics[width = 3 in]{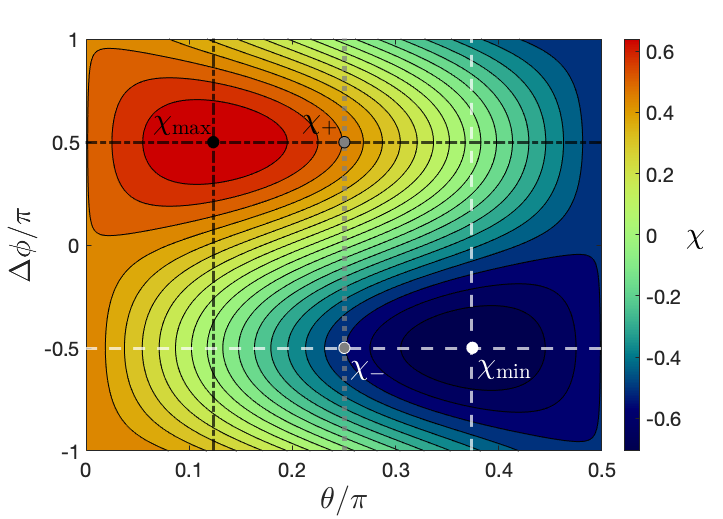}}\\
    \subfloat[]{\includegraphics[width = 3 in]{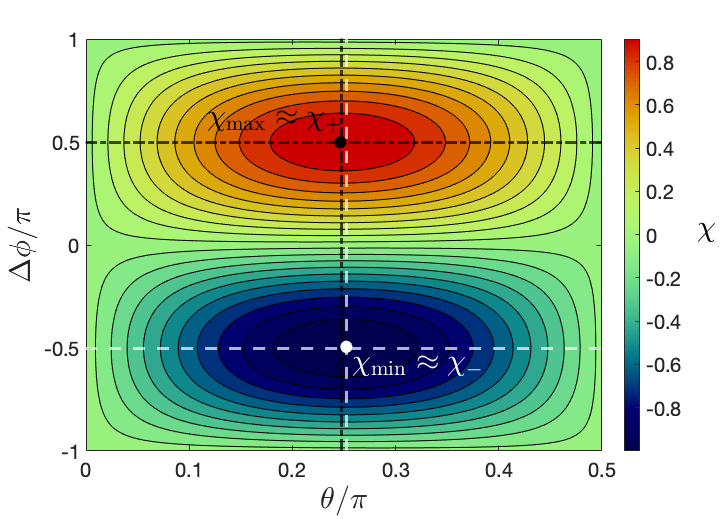}}
\caption{Directionality parameter $\chi$ as a function of the initial atomic state parameters,  for waveguide coupling efficiencies of (a) $\beta = 1$ (b) $\beta = 0.01$. The propagation phase is fixed to be $\phi= \pi/2$ and the atomic separation is considered to be in the non-retarded regime  $(\gamma T \ll 1)$. (a) For $\beta = 1$, the maximum directionality is obtained for $\theta=\pi/ 8$ and $\Delta\phi=\pm\pi/2$ with a maximum and minimum $\chi $ values of  $\chi_\mr{max}= -\chi_\mr{min} = 1/\sqrt{2}$. The gray points indicate directional emission obtained for initial states $ \ket{eg}\pm i \ket{ge}$ with a $\chi_\pm = \pm 1/2 $ \cite{Gheeraert20}. (b) For $\beta = 0.01$ the directional parameter can be as large as $ \chi_\mr{max} = -\chi_{\mr{min}} \approx 0.999$. }
    \label{Fig: chi}
\end{figure}

Fig.\,\ref{Fig: chi} shows the directionality of photon emission $\chi$ as a function of the parameters of the initial atomic state for  the optimum  directionality condition $\phi=\bkt{n+\frac{1}{2}}\pi$, for  two particular waveguide coupling efficiencies $(\beta = 1$ and $0.01)$.  Considering two orthogonal entangled  atomic states  $\ket{\Psi_{a (b)}}=\frac{1}{\sqrt{2}} \bkt{ \ket{eg} + e^{i\varphi_{a(b)}} \ket{ge}}$ with $\varphi_a - \varphi_b = \pi$, we obtain a  directional parameter value $\chi_{a(b)}=\sin\bkt{ \phi_R - \phi_L-\varphi_{a(b)} }/\bkt{1+\beta^{2}}$. It can be thus seen  that appropriately manipulating  the relative field propagation phase $(\phi_R - \phi_L)$ can allow one to distinguish any two orthogonal entangled states   based on the direction of emission, as illustrated by the points $\chi_{+}$ and $\chi_{-}$ on Fig.\,\ref{Fig: chi}. 

We note that the directional emission from an entangled state benefits from $\beta<1$ as can be seen from comparing Fig.\,\ref{Fig: chi} (a) and (b). This can be understood from the series expansions in Eqs. \eqref{eq:c12}, where the terms with odd powers of $\beta$ contribute to the directionality, while terms of order $\beta^2$ are detrimental. This contrasts with the standard case of neglecting field propagation, where $\beta<1$ does not change the qualitative behaviour of guided emission. In the limit $\beta\ll1$ \eqn{\label{Eq:chi1}
\chi(\Psi_0)\approx-\mathcal{C}\sin\phi\sin\Delta\phi,} where $\mathcal{C}=\sin2\theta$ is the concurrence that characterizes the entanglement of a pure state $\ket{\Psi_0}$. Thus, for small waveguide coupling efficiencies, the directionality $\chi$ could be a direct measure of the entanglement of the emitters. This also indicates that directional emission can be observed even in experiments with low coupling efficiency between emitters and waveguides.

We discuss the various parameter dependencies of the directionality below:

\begin{itemize}
    \item{Dependence on average propagation phase  $\bkt{\phi = \frac{\phi_R + \phi_L}{2}}$: Directionality comes from having constructive interference in one direction and destructive interference in the other direction.  Considering the phases of field propagation  $\phi_{R}=2 n\pi$ and $\phi_{L}=\left(2m+1\right)\pi$, gives $\phi=\bkt{n+\frac{1}{2}}\pi$. We see that this  maximizes the overall prefactor $\sin\phi$ and thus the directionality.
}
\item{Dependence on the relative atomic and field phases $\bkt{\Delta \phi}$: The interference effects between the atomic dipoles and the field  are represented by the $\sin\Delta\phi$ term, which maximizes directionality for   $\sin\Delta\phi\rightarrow1$.
This can be clearly seen from both Fig. \ref{Fig: chi}\,(a) and (b).}

\item{Dependence on waveguide coupling efficiency $(\beta)$ and initial atomic excitation amplitudes ($\theta$):  In the case of $\beta\rightarrow0$, there is nearly zero probability of multiple reflections, absorptions, and reemissions of the field in the waveguide, such that the field from one atom perfectly interferes (constructive or destructively) with the field from the other atom. For $\beta\rightarrow1$, the optimum directionality occurs when one atom radiates most of the field, while the second provides just enough field required for constructive interference (as seen from Fig.\ref{Fig: chi}\,(a)). This case corresponds to the optimum value $\theta=\pi/8$.
}
\end{itemize}

 Directionality can aid in sensing either the relative atomic phases or field propagation phases $\varphi$ \footnote{Here $\varphi $ could refer to either the atomic   phases $\cbkt{\phi_{A_1}, \phi_{A_2}}$, the field propagation phases   $\cbkt{\phi_{R},\phi_{L}} $, or the relative phases   $ \cbkt{\Delta \phi, \phi} $.}, all other experimental parameters being fixed.  In order to quantify this advantage we  define  $\mathcal{F}_D\bkt{\varphi}\equiv\sum_{i = L, R, \mr{out}} P_i \bkt{\partial\log\sbkt{P_i}/\partial\varphi}^2$ as the Fisher information that considers directionality. Here $P_i$ is the probability of emission into the decay channel $i$, spanning over  modes propagating to the left $(P_{\rm{L}})$,  right $(P_{\rm{R}})$, and out of the waveguide $(P_{\rm{out}}=1-(P_{\rm{R}}+P_{\rm{L}}))$. To compare with the case where one ignores directionality, we define the non-directional Fisher information $\mathcal{F}_{\rm{ND}}\bkt{\varphi}\equiv\sum_{i = \mr{tot}, \mr{out}} P_i \bkt{\partial\log\sbkt{P_i}/\partial\varphi}^2$, that considers only the total decay into and outside the waveguide with probabilities  $P_{\rm{tot}}$ and $P_{\rm{out}}$, respectively. It can be shown that $\mathcal{F}_{D}\bkt{\varphi}\geq\mathcal{F}_{ND}\bkt{\varphi}$ (see Appendix .\,\ref{Appen. D}), meaning that distinguishing the direction of propagation of the emitted photon helps to better estimate the general phase $\varphi$.

\section{Implementation in a circuit QED system}
\label{Appen. E}

\begin{figure}[ht]
    \centering
    \includegraphics[width = 3 in]{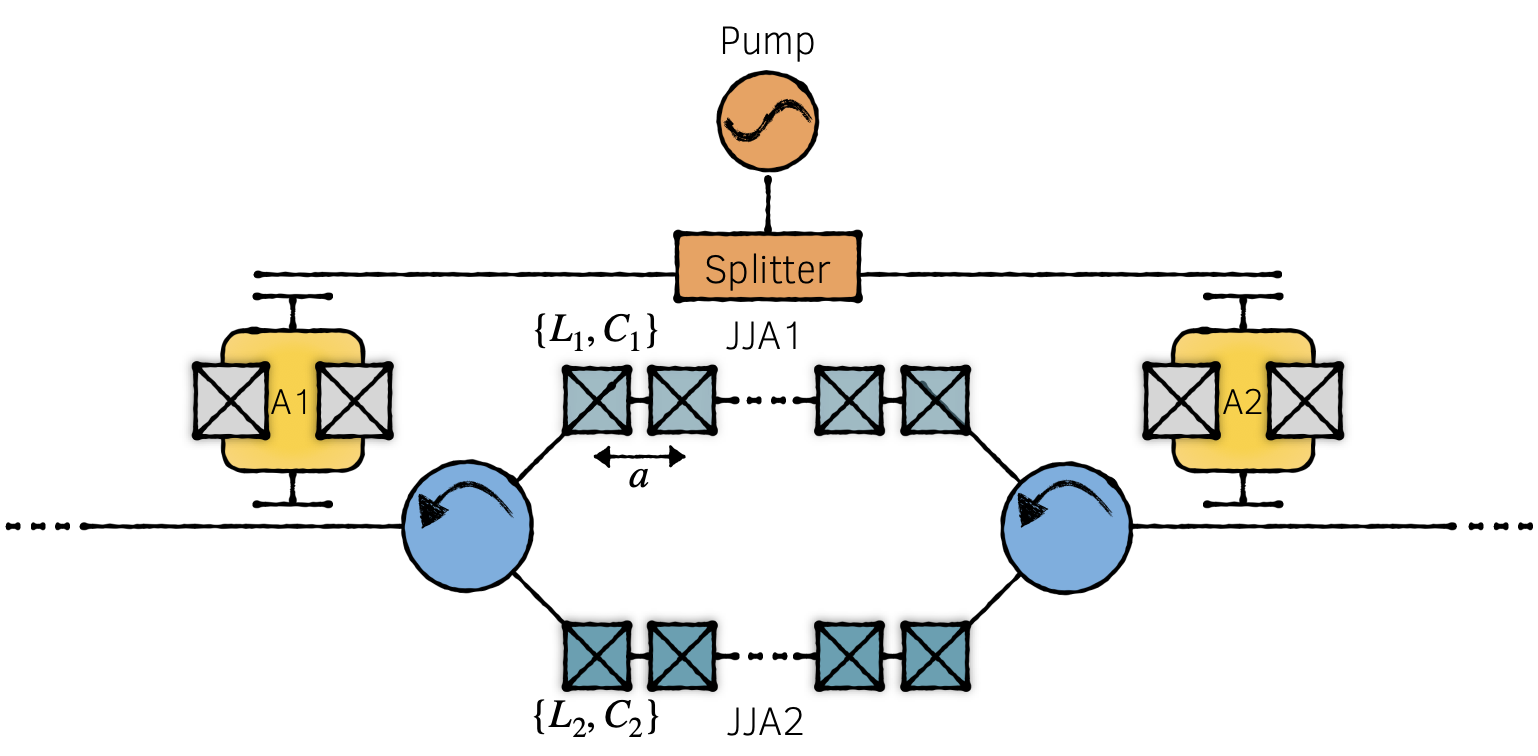}
    \caption{Schematic representation of a circuit QED implementation of the model. Two transmons (A1 and A2) coupled via circulators and Josephson junction (JJ) arrays (JJA1 and JJA2). The transmons are also coupled to each other via a control line that can allow for entangled state preparation by driving the qubits with an external pump field.}
    \label{Fig:cQED}
\end{figure}

The proposed system can be implemented with field circulators, that are readily available for fiber optics and an active element of research in superconducting circuits \cite{Kerckhoff2015,wang2021} and integrated photonics \cite{Wang15,Huang17}. These can be integrated into state-of-the art waveguide QED platforms \cite{sheremet2021}. 

We discuss a possible implementation of the present model in a circuit QED platform, as shown in Fig. \ref{Fig:cQED}.
 One can consider a system of two transmons  coupled via two separate Josephson junction arrays (JJAs), that allow for low-loss propagation of microwave fields with slow velocities \cite{JJA1}. We assume some typical parameter values for the proposed system as shown in Table\,\ref{parameters}.

\begin{table}[b]
\begin{center}
\begin{tabular}{|p {5 cm} |p {2 cm}|}
\hline
{Qubit resonance frequency $\omega_0/(2\pi)$}  &  5\,GHz  \\
\hline
{Decay rate $\gamma/(2\pi)$ } & 10\,MHz  \\
\hline
{Waveguide coupling efficiency $\beta$ } & 0.95\\
\hline
{Phase velocity in JJA1 $v_1/c$ } & 0.0033\\
\hline
{Phase velocity in JJA2 $v_2/c$ } & 0.0026\\
\hline
\end{tabular}
\caption{Parameter values for a superconducting circuit implementation of the model as depicted in Fig.\,\ref{Fig:cQED}.}
\label{parameters}
\end{center}
\end{table}
 The JJ arrays are considered to be made of $N\approx 2000$ junctions, and in the regime of relevant frequencies each junction can be modeled as a linear LC-oscillator, with inductance $L_{1(2)}$ and capacitance $C_{1(2)}$, capacitively coupled to the ground with a ground capacitance $C_g$. The inductance and capacitance values are assumed to be: $L_1 \approx1 $\,nH, $L_2 \approx1.58 $\,nH, $C_1\approx C_2 \approx1 $\,fF, $C_g \approx100 $\,fF \cite{JJA1,JJA2,JJA3}. The size of the unit cell in the JJ array is assumed to be $a = 10\,\mu$m. With the above set of parameters, and a  distance $ d\approx1.6$\,cm (such that $\gamma d/v\approx1$) between the emitters one can realize the parameter values considered.

\section{Summary and outlook}
\label{Sec: Summary and outlook}
We have proposed a system composed of two distant emitters coupled via a waveguide where the guided field experiences a direction-dependent propagation velocity (Fig.\,\ref{Fig:sch}). We show that in such a system the collective decay of the emitters can be non-simultaneous and, with an appropriate set of parameters, while one of the atoms can exhibit superradiance the other behaves subradiantly (Fig.\,\ref{fig:decays}(a)). {This suggests that collective decay can be described by local atom-photon interference  processes that lead to a mutually stimulated emission of the atoms (Eq. \,\eqref{eq:c12}).} The power radiated by such a pair of emitters can have a high directionality controlled by their phase relation and field propagation phase (Fig.\,\ref{fig:decays}(b)). Such directional emission is a rather general feature of collective delocalized systems (Eq.\,\eqref{Eq:chi}). We analyze the directionality of emission as a function of various parameters, characterizing the optimal conditions for directional emission (Fig.\,\ref{Fig: chi}).  Our results suggest that such directional emission can also be observed for waveguides with low coupling efficiencies. We further remark that an analog of the phenomena described here can also be observed in classical systems \cite{Lama72}. Nonetheless, in the proposed model, the directionallity of the coupling  aids the detection of entanglement (Eq.\,\eqref{Eq:chi1}) and  helps distinguish between the symmetric and asymmetric entangled states of two emitters (Fig.\,\ref{Fig: chi}). We finally propose a possible implementation of the scheme in a superconducting circuit platform in Sec.~\ref{Appen. E}.

While on the one hand our results show that directional emission could be used for state tomography and measuring entanglement, on the other hand one can prepare the emitters in an entangled state by driving them through the waveguide. This can be thought as the time reversal process of collective directional emission \cite{Yang2021, Zens2021,Sinha20}. The directional emission and state preparation protocols can allow for efficient and controllable routing of quantum information in quantum networks \cite{Guimond20, Gheeraert20, Kimble08, Schoelkopf08}.

The phenomena described in this work can be extended to study directional emission from collective many-body quantum states, with {the presented system as a fundamental unit along a chain of emitters coupled to an \textit{anisotachyic} bath.}  Additionally, for strongly driven systems, the effects of atomic nonlinearity become relevant \cite{Hughes2020, Hughes2021}. It has been shown, for example, that nonlinearity can assist in directional emission \cite{Kaplan81, Kaplan82, Hamann,  Guimond20}. It would therefore be pertinent to analyze and optimize the directionality over a broader set of parameters including general atomic states, field propagation phases in nonlinear systems and anisotachy. 

Anisotachy in waveguide QED platforms could offer new ways to manipulate light-matter interactions. In particular, we show here that it can be used to couple delocalized correlated state of two emitters to a specific direction of collective radiation. This effect expands the toolbox for quantum optics applications while enriching our understanding of waveguide QED systems.

\section{Acknowledgments} We are grateful to Pierre Meystre and Alejandro Gonz\'{a}lez-Tudela for insightful comments on the manuscript. This work was supported in part by CONICYT-PAI grant 77190033, FONDECYT grant N$^{\circ}$ 11200192 from Chile, and grant No. UNAM-DGAPA-PAPIIT IG101421 from Mexico.

\appendix

\begin{widetext}
\section{Derivation of the equations of motion}
\label{Appen A}

We can write the coupled  emitter-field equations of motion from Eqs.~\eqref{Eq:psit} and \eqref{eq:c12} as: 

\eqn{\label{ca1}
\dot {c}_{\rm{R}}  \bkt{\omega, t}&= -i \sum_{m = 1,2} c_m \bkt{t} \bkt{g\bkt{\omega}}^\ast e^{-i \omega x_m /v_{\rm{R}}} e^{i\bkt{ \omega - \omega_0 } t }, \\
\label{cb1}
\dot {c}_{\rm{L}}  \bkt{\omega, t}&=  -i \sum_{m = 1,2} c_m \bkt{t}\bkt{ g\bkt{\omega}}^\ast e^{i \omega x_m /v_{\rm{L}}} e^{i\bkt{ \omega - \omega_0 } t },\\
\label{cm1}
\dot {c}_m  \bkt{t}&= -i \int_0 ^\infty \dd\omega \,g\bkt{\omega} e^{-i \bkt{ \omega - \omega_ 0} t}\sbkt{ c_{\rm{R}} \bkt{ \omega, t} e^{i \omega x_m/v_{\rm{R}}} +c_{\rm{L}} \bkt{ \omega, t} e^{-i \omega x_m/v_{\rm{L}}} }.
}

Formal integration of \eqref{ca1} and \eqref{cb1} yields
\eqn{
c_{\rm{R}}\bkt{\omega, t} &=-i \int_0 ^t \dd\tau\, \sum_{m = 1, 2} g^\ast \bkt{\omega} c_m\bkt{\tau} e^{-i \omega x_m/v_{\rm{R}}} e^{i \bkt{\omega - \omega_0 }\tau },\\
c_{\rm{L}}\bkt{\omega, t} &= -i \int_0 ^t \dd\tau\, \sum_{m = 1, 2} g^\ast \bkt{\omega} c_m\bkt{\tau} e^{i \omega x_m /v_{\rm{L}}} e^{i \bkt{\omega - \omega_0 }\tau }.
}

Substituting the above in Eq.\eqref{cm1}, we can rewrite the atomic  equation  as follows

\eqn{
\dot{c}_m\bkt{t}  =& - \int_0 ^\infty\dd\omega \, \abs{g\bkt{\omega}}^2 \int_0 ^t\dd\tau\sum_{n = 1, 2} c_n \bkt{\tau}e^{-i \bkt{\omega - \omega_0 }\bkt{t - \tau} }\sbkt{ e^{ i \omega \bkt{x_m - x_n} /v_{\rm{R}}} + e^{ -i \omega \bkt{x_m - x_n} /v_{\rm{L}}}}.
}

We now define the field correlation function $F\bkt{s}=\int_0 ^\infty\dd\omega \abs{g\bkt{\omega}}^2e^{-i \bkt{\omega - \omega_0 }s}$, to obtain
\eqn{
\dot{c}_m\bkt{t}  =& - \int_0 ^t\dd\tau\sbkt{2c_m\bkt{\tau} F\bkt{t-\tau} +c_n \bkt{\tau}\cbkt{ e^{ i \omega_0 T^{mn}_{\rm{R}}}F\bkt{t-\tau-T^{mn}_{\rm{R}}}+ e^{ -i \omega_0 T^{mn}_{\rm{L}}}F\bkt{t-\tau+T^{mn}_{\rm{L}}}}},
}

where $T^{mn}_{\rm{R},\rm{L}}=\bkt{x_m - x_n} /v_{\rm{R},\rm{L}}$ is the direction dependent delay time for the light propagating between the emitters.

In the standard Markov approximation $F\bkt{\tau}=2\pi\abs{g\bkt{\omega_0}}^2\delta(\tau) $, though more generally $F\bkt{\tau}$ is a narrow distribution symmetric around $s=0$. We assume  that the temporal width $\sigma$ of such distribution is narrower than the delay time between the emitters ($\sigma<|T_{L,R}^{mn}|$), 

\eqn{
\dot{c}_m\bkt{t}  =& - 2\int_0 ^t\dd \tau\, c_m\bkt{t-\tau} F\bkt{\tau}-e^{i \omega_0 T^{mn}}\int_{-T^{mn}} ^{t-T^{mn}}\dd \tau\, c_n \bkt{t-\tau-T^{mn}} F\bkt{\tau},
}
where  $T^{12} = T^{12}_{\rm{L}}=T_{\rm{L}}$ or $T^{21} = T^{21}_{\rm{R}}=T_{\rm{R}}$. If $\sigma$ is small enough we can assume that the amplitude of the coefficients  does not vary significantly over the region where $F\bkt{\tau}$ is non-zero, such that $c_m\bkt{t-\tau}\approx c_m\bkt{t}$. Thus given that $F\bkt{\tau}$ is symmetric,  centered around $\tau=0$ and narrower than $T^{mn}$ we have 

\eqn{
\dot{c}_m\bkt{t}  \approx & - 2c_m\bkt{t}\int_0 ^{\infty}\dd s\,  F\bkt{s}-c_n \bkt{t-T^{mn}}\Theta\bkt{t-T^{mn}}e^{i \omega_0 T^{mn}}\int_{-\infty} ^{\infty}\dd \tau\,  F\bkt{\tau}.
\label{cmdot}
}

The term $F\bkt{\tau}$ is a complex function with the real and imaginary part being even and odd functions respectively.
We define
\eqn{\frac{\gamma}{2}=& 2 \re\sbkt{\int_{0} ^{\infty}\dd \tau\,F\bkt{\tau}}= \re\sbkt{\int_{-\infty} ^{\infty}\dd \tau\,F\bkt{\tau}},\\
\Delta_{L}=& \im\sbkt{\int_{0} ^{\infty}\dd \tau\,F\bkt{\tau}},
}
where $\Delta_{L}$ is the Lamb shift, which we include as a part of the emitters renormalized  resonance frequency $\omega_0$.
 
Introducing  a phenomenological cross-coupling efficiency $\beta$ between the emitters ($0\leq\beta\leq 1$), one can simplify Eq.\,\eqref{cmdot} to obtain the atomic equations of motion (Eq.~\eqref{eq:c12}).

\section{Atomic dynamics}
\label{Appen. B}

\subsection{Lambert W-function solution}

Taking the Laplace transform of Eq.~\eqref{eq:c12}, one gets
\eqn{
s \tilde{c}_1\bkt{s}-c_{1}(0)  =& - \frac{\gamma}{2} \sbkt{\tilde{c}_1\bkt{s}+\beta \tilde{c}_2\bkt{s} e^{ -sT_{\mr{L}}}e^{i \phi_{\mr{L}}}},\\
s \tilde{c}_2\bkt{s}-c_2(0)  =& - \frac{\gamma}{2} \sbkt{\tilde{c}_2\bkt{s}+\beta \tilde{c}_1\bkt{s} e^{ -sT_{\mr{R}}}e^{i \phi_{\mr{R}}}},
}
which can be solved to obtain the Laplace coefficients pertaining to the two emitters as follows
\eqn{
\tilde{c}_1\bkt{s}&=\frac{c_{1}(0)\bkt{s+\frac{\gamma}{2}}-c_2(0)\beta\frac{\gamma}{2}e^{-sT_{\rm{L}}}e^{i \phi_{\rm{L}}}}{\bkt{s+\frac{\gamma}{2}}^2-\bkt{\beta\frac{\gamma}{2}e^{-sT}e^{i \phi}}^2}\label{eq:c1Laplace},\\
\tilde{c}_2\bkt{s}&=\frac{c_2(0)\bkt{s+\frac{\gamma}{2}}-c_{1}(0)\beta\frac{\gamma}{2}e^{-sT_{\rm{R}}}e^{i \phi_{\rm{R}}}}{\bkt{s+\frac{\gamma}{2}}^2-\bkt{\beta\frac{\gamma}{2}e^{-sT}e^{i \phi}}^2}\label{eq:c2Laplace}.
}

The poles of the above Laplace coefficients are given by
\eqn{
s^\pm_{ n}=-\frac{\gamma}{2}+\frac{1}{T} W_{n}\sbkt{\mp \beta\frac{ \gamma T}{2} e^{\gamma T/2}e^{i \phi}},
}
where $W_n$ is the $n^\mr{th}$ branch of the Lambert W-function \cite{Corless96}.

We can thus rewrite the Eq. (\ref{eq:c1Laplace}) and (\ref{eq:c2Laplace}) as
\eqn{\label{cms}
\tilde{c}_m\bkt{s}&=\sum_\pm\sum_{n=-\infty}^{\infty}\frac{\alpha_{ n,m}^{\pm}}{s-s_{ n}^\pm},
}
where the coefficients $\alpha_{ n,m}^{\pm}$ are obtained as
\eqn{\alpha_{ n,m}^{\pm}=\lim_{s\rightarrow s_{ n}^\pm}\tilde{c}_m\bkt{s}\bkt{s-s_{n}^\pm }.
}

Thus taking the inverse Laplace transform of Eq.\,\eqref{cms}, we  get
\eqn{\label{Eq:cmtw}
c_m\bkt{t}=\sum_{\sigma = +,-}\sum_{n=-\infty}^{\infty}\alpha_{ n, m}^{\sigma}e^{-\gamma_{ n}^\sigma t},
}

where 
\eqn{
\gamma_{ n}^\pm&=\frac{\gamma}{2}-\frac{1}{T} W_{n}\bkt{\mp \frac{\beta\gamma T}{2} e^{\gamma T/2}e^{i \phi}},\\
\alpha_{ n,1}^{\pm}&=\frac{1}{2}\frac{c_{1}(0)\pm c_2(0) e^{i\bkt{\phi_L - \phi_R }/2}e^{\bkt{T_L - T_R } \gamma^\pm_{ n}/2}}{1+W_n\bkt{\mp \frac{\beta\gamma T}{2} e^{\gamma T/2}e^{i \phi}}},\\
\alpha_{ n,2}^{\pm}&=\frac{1}{2}\frac{c_2(0)\pm c_{1}(0) e^{-i\bkt{\phi_L - \phi_R }/2}e^{-\bkt{T_L - T_R }\gamma^\pm_{ n}/2}}{1+W_n\bkt{\mp \frac{\beta\gamma T}{2} e^{\gamma T/2}e^{i \phi}}}.
}

\subsection{Series expansion solution}

An alternative way of expressing the  atomic excitation amplitudes  as the inverse Laplace transform of  Eq.\,\eqref{eq:c1Laplace} and \eqref{eq:c2Laplace} in terms of a series solution is as follows \cite{Milonni74}:

\eqn{
c_{1}\bkt{t} =& \frac{1}{2\pi i } \lim_{\epsilon\rightarrow0}\int_{-i \infty + \epsilon}^{+i \infty + \epsilon} \dd s \frac{c_{1}(0)\bkt{s + \frac{\gamma}{2}} - c_2(0) \frac{\beta \gamma}{2}e^{- s T_L }e^{i \phi_L}}{\bkt{s + \frac{\gamma}{2}}^2} \sbkt{\sum_{n = 0}^\infty  \bkt{ \frac{\beta \gamma/2 e^{- sT} e^{i \phi}}{s + \gamma/2}}^{2n}},\\
= &c_{1}(0)\underbrace{  \frac{1}{2\pi i }\int \dd s \sbkt{ \frac{1}{s + \gamma/2} \cbkt{\sum_{n = 0}^\infty  \bkt{ \frac{\beta \gamma/2 e^{- sT} e^{i \phi}}{s + \gamma/2}}^{2n}}}}_{\mr{(Ia)}} \non\\
&-c_2(0) \underbrace{ \frac{1}{2\pi i }\int \dd s \sbkt{ \frac{\beta \gamma}{2} \frac{ e^{- s T_L } e^{ i \phi_L }}{ \bkt{s + \gamma/2}^2}\cbkt{\sum_{n = 0}^\infty  \bkt{ \frac{\beta \gamma/2 e^{- sT} e^{i \phi}}{s + \gamma/2}}^{2n}}} }_ {\mr{(IIa)}}.
\label{Eq:c1t}
}
\eqn{
c_{2}\bkt{t} =& \frac{1}{2\pi i } \lim_{\epsilon\rightarrow0}\int_{-i \infty + \epsilon}^{+i \infty + \epsilon} \dd s \frac{c_2(0)\bkt{s + \frac{\gamma}{2}} -c_{1}(0) \frac{\beta \gamma}{2}e^{- s T_R }e^{i \phi_R}}{\bkt{s + \frac{\gamma}{2}}^2} \sbkt{\sum_{n = 0}^\infty  \bkt{ \frac{\beta \gamma/2 e^{- sT} e^{i \phi}}{s + \gamma/2}}^{2n}},\\
= &c_2(0)\underbrace{  \frac{1}{2\pi i }\int \dd s \sbkt{ \frac{1}{s + \gamma/2} \cbkt{\sum_{n = 0}^\infty  \bkt{ \frac{\beta \gamma/2 e^{- sT} e^{i \phi}}{s + \gamma/2}}^{2n}}}}_{\mr{(Ib)}}\non\\
&- c_{1}(0)\underbrace{ \frac{1}{2\pi i }\int \dd s \sbkt{ \frac{\beta \gamma}{2} \frac{ e^{- s T_R } e^{ i \phi_R }}{ \bkt{s + \gamma/2}^2}\cbkt{\sum_{n = 0}^\infty  \bkt{ \frac{\beta \gamma/2 e^{- sT} e^{i \phi}}{s + \gamma/2}}^{2n}}} }_ {\mr{(IIb)}}.
\label{Eq:c2t}
}

We identify the terms (Ia) = (Ib) $\equiv $ (I) as corresponding to the round trip times (even number) of the field between the atoms and the terms (IIa) and (IIb) (not necessarily equal to each other) as the terms coming from odd number of trips between the atoms.  Simplifying each of the above terms:

\eqn{
\mr{(I)}=& \frac{1}{2\pi i }\int \dd s \sbkt{ \frac{1}{s + \gamma/2} \cbkt{\sum_{n = 0}^\infty  \bkt{ \frac{\beta \gamma/2 e^{- sT} e^{i \phi}}{s + \gamma/2}}^{2n}}},\\
=&  \sum_{n}\frac{\bkt{\beta \gamma e^{i \phi}/2}^{2n}}{(2n)!}  \bkt{t - 2n T }^{2n} e^{- \gamma /2 \bkt{ t - 2nT}}\Theta\bkt{t - 2 n T},\\
(\mr{IIa} )=& \frac{1}{2\pi i }\int \dd s \sbkt{ \frac{\beta \gamma}{2} \frac{ e^{- s T_L } e^{ i \phi_L }}{ \bkt{s + \gamma/2}^2}\cbkt{\sum_{n = 0}^\infty  \bkt{ \frac{\beta \gamma/2 e^{- sT} e^{i \phi}}{s + \gamma/2}}^{2n}}} ,\\
=& \sum_{n}\frac{\bkt{\beta \gamma /2}^{2n+1}}{(2n+1)!} e^{i 2n\phi+ i \phi_L}  \bkt{t - 2n T - T_L }^{2n+1} e^{- \gamma /2 \bkt{ t - 2nT- T_L}}\Theta\bkt{t - 2 n T- T_L},\\
(\mr{IIb} )=& \frac{1}{2\pi i }\int \dd s \sbkt{ \frac{\beta \gamma}{2} \frac{ e^{- s T_R } e^{ i \phi_R }}{ \bkt{s + \gamma/2}^2}\cbkt{\sum_{n = 0}^\infty  \bkt{ \frac{\beta \gamma/2 e^{- sT} e^{i \phi}}{s + \gamma/2}}^{2n}}}, \\
=& \sum_{n}\frac{\bkt{\beta \gamma /2}^{2n+1}}{(2n+1)!} e^{i 2n\phi+ i \phi_R}  \bkt{t - 2n T - T_R }^{2n+1} e^{- \gamma /2 \bkt{ t - 2nT- T_R}}\Theta\bkt{t - 2 n T- T_R}.
}

We substitute the above in Eqs.\,\eqref{Eq:c1t} and \eqref{Eq:c2t} to obtain the dynamics of general initial states given by Eq.\,\eqref{eq:c12}.

\begin{figure}
    \centering
    \includegraphics[width = 5 in]{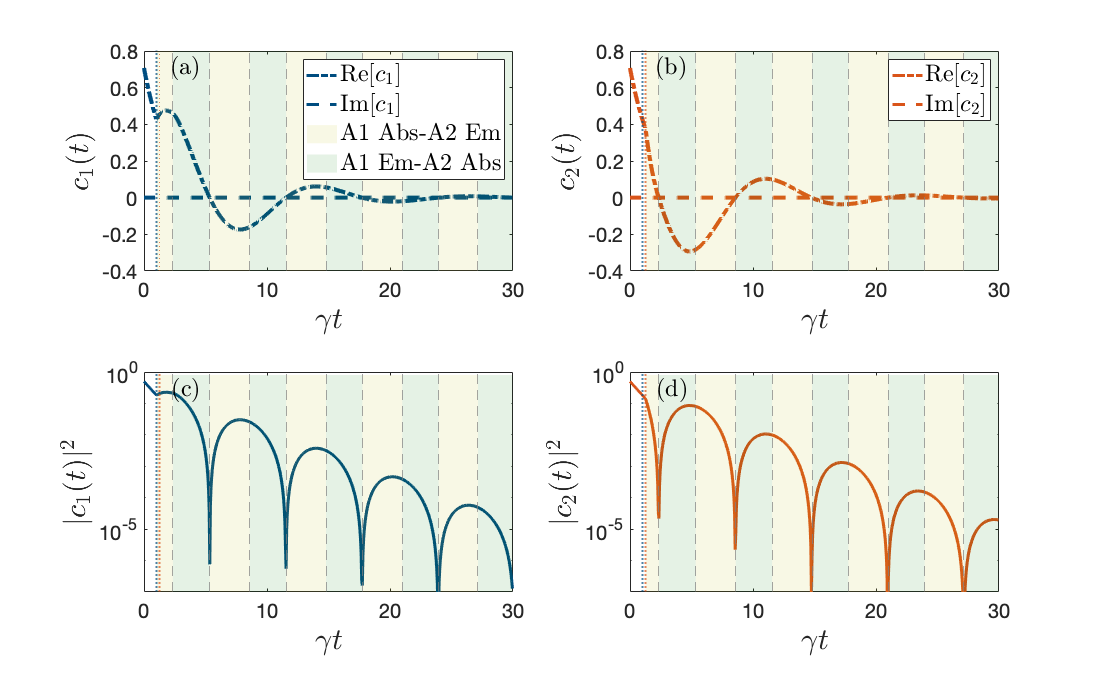}
    \caption{Dynamics of atomic coefficients (a) $c_1(t) $ and (b) $c_2(t)$, and atomic populations (c) $\abs{c_1\bkt{t}}^2 $ and (d) $\abs{c_2\bkt{t}}^2 $. Whenever the atomic coefficients change sign ($c_{1,2}(t)=0$) the atomic populations reverse their absorption and emission  behavior, as indicated by the yellow and green regions corresponding to atom A1 absorbing-A2 emitting and A1 emitting-A2 absorbing, respectively.}
    \label{Fig:longtime}
\end{figure}

We plot the atomic dynamics as a function of time in Fig.\,\ref{Fig:longtime}. It can be seen that when the sign of the atomic coefficients $c_{1,2}$ changes, so does the sign of its electric dipole moment that drives the field, causing the atoms to switch from absorbing to emitting, or vice versa.

\section{Intensity dynamics}
\label{Appen. C}

The intensity of the field emitted by the atoms as a function of position and time can be evaluated as $I\bkt{x,t} = \frac{\epsilon_0 c} {2}\avg{\Psi\bkt{t}\vert\hat{E}^\dagger\bkt{x,t} \hat E\bkt{x,t}\vert\Psi\bkt{t}} $, where $\hat{E}\bkt{x,t} = \int_0^\infty  \dd \omega\,\mc{E}_\omega \sbkt{\hat{a}_L \bkt{\omega}e^{-i k_L x} + \hat{a}_R \bkt{\omega}e^{i k_R x}}e^{-i \omega t}$ is the electric field operator at position $x$ and time $t$. More explicitly, we obtain 

\eqn{
I(x,t)  =& \frac{\epsilon_0 c \abs{\mc{E}_0 }^2}{2} \bra{\Psi\bkt{t}}\sbkt{\int \dd \omega_1  \cbkt{\hat{a}_L^\dagger \bkt{\omega_1}e^{i k_{1L} x} + \hat{a}_R^\dagger \bkt{\omega_1}e^{-i k_{1R }x}}e^{i \omega_1 t}\right.\non\\
&\left.\int \dd \omega_2\cbkt{\hat{a}_L \bkt{\omega_2}e^{-i k_{2L} x} + \hat{a}_R \bkt{\omega_2}e^{i k_{2R} x}}e^{-i \omega_2 t}} \ket{\Psi\bkt{t}},\\
 = &\frac{\epsilon_0 c \abs{\mc{E}_0 }^2}{2} \bra{\Psi\bkt{t}} \int \dd \omega_1  \int \dd \omega_2  \sbkt{ e^{i \bkt{ k_{1L} - k_{2L}} x} c_L^\ast \bkt{ \omega_1, t} c_L\bkt{ \omega_2, t} + e^{-i \bkt{ k_{1R} - k_{2R}} x} c_R^\ast \bkt{ \omega_1, t} c_R\bkt{ \omega_2, t} \right.\non\\
 &\left. + e^{-i \bkt{ k_{1R} + k_{2L}} x} c_R^\ast \bkt{ \omega_1, t} c_L\bkt{ \omega_2, t} + e^{i \bkt{ k_{1L} + k_{2R}} x} c_L^\ast \bkt{ \omega_1, t} c_R\bkt{ \omega_2, t}}e^{i \bkt{\omega_1 -\omega_2}t}\ket{\Psi\bkt{t}},\\
 =&\frac{\epsilon_0 c \abs{\mc{E}_0 }^2}{2} \bra{\Psi\bkt{t}} \abs{ \int \dd\omega \sbkt{c_L \bkt{\omega,t}e^{-i\omega x/v_L} + c_R\bkt{\omega,t} e^{i\omega x/v_R}}e^{-i\omega t}}^2\ket{\Psi\bkt{t}},\\
 =& \frac{\epsilon_0 c \abs{\mc{E}_0 }^2\gamma \beta}{4\pi }\abs{ \int \dd\omega e^{-i \omega t} \sbkt{\int_0 ^t \dd \tau\cbkt{ c_{1}\bkt{\tau} e^{i \omega (-x+x_1)/v_L} + c_{2}\bkt{\tau} e^{i \omega (-x+ x_2)/v_L}\right.\right.\right.\non\\
 &\left.\left.\left.+  c_{1}\bkt{\tau} e^{-i \omega (-x+x_1)/v_R} + c_{2}\bkt{\tau} e^{-i \omega (-x+x_2)/v_R}}e^{i \bkt{\omega- \omega_0 }\tau}}}^2,}
 
 where we have used Eqs.\,\eqref{ca1} and \eqref{cb1} to substitute the field amplitudes in terms of the atomic excitation amplitudes. Using the W-function solution for the atomic coefficients (Eq.\,\eqref{Eq:cmtw}) and performing the integrals over time and frequency, we obtain
 
 \eqn{
I/I_0 = &\abs{ \sum_{\sigma = +,-}\sum_{n = -\infty}^\infty \sbkt{\alpha_{n,1}^\sigma  e^{-i \bkt{\omega_0 - i \gamma_n^\sigma}\bkt{t+\bkt{x - x_1}/v_L }}\cbkt{ \Theta \sbkt{t+\bkt{x - x_1}/v_L} -\Theta \sbkt{\bkt{x - x_1}/v_L} }\right.\right.\non\\
 &\left.\left. +\alpha_{n,2}^\sigma  e^{-i \bkt{\omega_0 - i \gamma_n^\sigma}\bkt{t+\bkt{x - x_2}/v_L }}\cbkt{ \Theta \sbkt{t+\bkt{x - x_2}/v_L} -\Theta \sbkt{\bkt{x - x_2}/v_L} }\right.\right.\non\\
 &\left.\left. +\alpha_{n,1}^\sigma  e^{-i \bkt{\omega_0 - i \gamma_n^\sigma}\bkt{t-\bkt{x - x_1}/v_R }}\cbkt{ \Theta \sbkt{t-\bkt{x - x_1}/v_R} -\Theta \sbkt{-\bkt{x - x_1}/v_R} }\right.\right.\non\\
 &\left.\left. +\alpha_{n,2}^\sigma  e^{-i \bkt{\omega_0 - i \gamma_n^\sigma}\bkt{t-\bkt{x - x_2}/v_R }}\cbkt{ \Theta \sbkt{t-\bkt{x - x_2}/v_R} -\Theta \sbkt{-\bkt{x - x_2}/v_R} } }}^2.}
 We can rewrite the above in terms of the atomic excitation amplitudes using Eq.\,\eqref{Eq:cmtw} as
 \eqn{
I/I_0   = &\abs{  \sbkt{c_1\bkt{t+\bkt{x +d/2}/v_L } e^{-i\omega_0 \bkt{x +d/2}/v_L } \cbkt{ \Theta \sbkt{t+\bkt{x +d/2}/v_L} -\Theta \sbkt{\bkt{x +d/2}/v_L} }\right.\right.\non\\
 &\left.\left. +c_2\bkt{ t+\bkt{x - d/2}/v_L}  e^{-i \omega_0 \bkt{x - d/2}/v_L }\cbkt{ \Theta \sbkt{t+\bkt{x -d/2}/v_L} -\Theta \sbkt{\bkt{x - d/2}/v_L} }\right.\right.\non\\
 &\left.\left. +c_1\bkt{t-\bkt{x +d/2}/v_R} e^{i \omega_0 \bkt{x +d/2}/v_R }\cbkt{ \Theta \sbkt{t-\bkt{x +d/2}/v_R} -\Theta \sbkt{-\bkt{x +d/2}/v_R} }\right.\right.\non\\
 &\left.\left. +c_2\bkt{t-\bkt{x - d/2}/v_R}  e^{i \omega_0 \bkt{x - d/2}/v_R }\cbkt{ \Theta \sbkt{t-\bkt{x - d/2}/v_R} -\Theta \sbkt{-\bkt{x - d/2}/v_R} } }}^2,
}

which corresponds to Eq.~\eqref{Eq:chi}.

\section{Directional Emission }
\label{Appen. D}

Let us consider the dynamics for the atomic coefficients given by Eq.~\eqref{eq:int}. In the limit $\gamma T\ll 1$, neglecting the delay but keeping the propagation phases, we obtain:

\begin{align}
c_{1}(t)= & \left[c_{1}\bkt{0}\sum_{n=0}^\infty\frac{\left(\beta\frac{\gamma}{2}te^{i\phi}\right)^{2n}}{2n!}-c_{2}\bkt{0}e^{i(\phi_{L}-\phi)}\sum_{n=0}^\infty\frac{\left(\beta\frac{\gamma}{2}te^{i\phi}\right)^{2n+1}}{(2n+1)!}\right]e^{-\frac{\gamma}{2}t}\Theta(t),\\
c_{2}(t)= & \left[c_{2}\bkt{0}\sum_{n =0 }^\infty\frac{\left(\beta\frac{\gamma}{2}te^{i\phi}\right)^{2n}}{2n!}-c_{1}\bkt{0}e^{i(\phi_{R}-\phi)}\sum_{n=0}^\infty\frac{\left(\beta\frac{\gamma}{2}te^{i\phi}\right)^{2n+1}}{(2n+1)!}\right]e^{-\frac{\gamma}{2}t}\Theta(t).
\end{align}

These series converges to
\begin{align}
c_{1}(t)= & \left[c_{1}\bkt{0}\cosh\left\{ \beta\frac{\gamma}{2}te^{i\phi}\right\} -c_{2}\bkt{0}e^{i(\phi_{L}-\phi)}\sinh\left\{ \beta\frac{\gamma}{2}te^{i\phi}\right\} \right]e^{-\frac{\gamma}{2}t}\Theta(t),\\
c_{2}(t)= & \left[c_{2}\bkt{0}\cosh\left\{ \beta\frac{\gamma}{2}te^{i\phi}\right\} -c_{1}\bkt{0}e^{i(\phi_{R}-\phi)}\sinh\left\{ \beta\frac{\gamma}{2}te^{i\phi}\right\} \right]e^{-\frac{\gamma}{2}t}\Theta(t).
\end{align}

We  now  consider the field coefficients in the steady state $(t\rightarrow\infty)$,
which  can be simplified to:
\begin{align}
&c_{R}(\omega,t\rightarrow\infty)= \non\\
&-ig^{\ast}(\omega)e^{-i\frac{\phi_{R}}{2}}\left[\left(c_{1}\bkt{0}e^{i\phi_{R}}+c_{2}\bkt{0}\right)\frac{(\frac{\gamma}{2}-i\Delta)}{(\frac{\gamma}{2}-i\Delta)^{2}-\left(\beta\frac{\gamma}{2}e^{i\phi}\right)^{2}}-\left(c_{2}\bkt{0}e^{2i\phi}+c_{1}\bkt{0}e^{i\phi_{R}}\right)\frac{\beta\frac{\gamma}{2}}{(\frac{\gamma}{2}-i\Delta)^{2}-\left(\beta\frac{\gamma}{2}e^{i\phi}\right)^{2}}\right],\\
&c_{L}(\omega,t\rightarrow\infty)= \non\\
&-ig^{\ast}(\omega)e^{-i\frac{\phi_{L}}{2}}\left[\left(c_{1}\bkt{0}+c_{2}\bkt{0}e^{i\phi_{L}}\right)\frac{(\frac{\gamma}{2}-i\Delta)}{(\frac{\gamma}{2}-i\Delta)^{2}-\left(\beta\frac{\gamma}{2}e^{i\phi}\right)^{2}}-\left(c_{2}\bkt{0}e^{i\phi_{L}}+c_{1}\bkt{0}e^{2i\phi}\right)\frac{\beta\frac{\gamma}{2}}{(\frac{\gamma}{2}-i\Delta)^{2}-\left(\beta\frac{\gamma}{2}e^{i\phi}\right)^{2}}\right].
\end{align}

The probability of emitting the photon to the right (left) is thus given by
\begin{equation}
    P_{R(L)}=\int_{0}^{\infty}d\omega\left|c_{R(L)}(\omega,t\rightarrow\infty)\right|^{2}.
\end{equation}

We parametrize the initial atomic coefficients as $c_{1}\bkt{0}=e^{i\phi_{A1}}\cos\theta$
and $c_{2}\bkt{0}=e^{i\phi_{A2}}\sin\theta$,  to obtain the right and left  emission probabilities as follows:

\begin{align}
\label{PR}
&P_{R} =\non\\
&\left|g(\omega_{0})\right|^{2}\int_{0}^{\infty}d\omega\left|\left(\sin\theta+\cos\theta e^{i\Delta\phi}e^{i\phi}\right)\frac{(\frac{\gamma}{2}-i\Delta)}{(\frac{\gamma}{2}-i\Delta)^{2}-\left(\beta\frac{\gamma}{2}e^{i\phi}\right)^{2}}-\left(\sin\theta+\cos\theta e^{i\Delta\phi}e^{-i\phi}\right)e^{2i\phi}\frac{\beta\frac{\gamma}{2}}{(\frac{\gamma}{2}-i\Delta)^{2}-\left(\beta\frac{\gamma}{2}e^{i\phi}\right)^{2}}\right|^{2},\\
\label{PL}
&P_{L}=\non\\
&\left|g(\omega_{0})\right|^{2}\int_{0}^{\infty}d\omega\left|\left(\sin\theta+\cos\theta e^{i\Delta\phi}e^{-i\phi}\right)\frac{(\frac{\gamma}{2}-i\Delta)}{(\frac{\gamma}{2}-i\Delta)^{2}-\left(\beta\frac{\gamma}{2}e^{i\phi}\right)^{2}}-\left(\sin\theta+\cos\theta e^{i\Delta\phi}e^{i\phi}\right)\frac{\beta\frac{\gamma}{2}}{(\frac{\gamma}{2}-i\Delta)^{2}-\left(\beta\frac{\gamma}{2}e^{i\phi}\right)^{2}}\right|^{2}.
\end{align}

We consider the integral:
\begin{align}
I_{0} & =\int_{0}^{\infty}d\omega\left|\frac{T_{1}(\frac{\gamma}{2}-i\Delta)-T_{2}\beta\frac{\gamma}{2}}{(\frac{\gamma}{2}-i\Delta)^{2}-\left(\beta\frac{\gamma}{2}e^{i\phi}\right)^{2}}\right|^{2},\\
& =\left|T_{1}\right|^{2}\int_{0}^{\infty}d\omega\frac{\Delta^{2}}{\left|(\frac{\gamma}{2}-i\Delta)^{2}-\left(\beta\frac{\gamma}{2}e^{i\phi}\right)^{2}\right|^{2}}-2\im\sbkt{ T_{1}T_{2}^{*}} \beta\frac{\gamma}{2}\int_{0}^{\infty}d\omega\frac{\Delta}{\left|(\frac{\gamma}{2}-i\Delta)^{2}-\left(\beta\frac{\gamma}{2}e^{i\phi}\right)^{2}\right|^{2}}\nonumber\\
 & +\left|T_{1}-T_{2}\beta\right|^{2}\left(\frac{\gamma}{2}\right)^{2}\int_{0}^{\infty}d\omega\frac{1}{\left|(\frac{\gamma}{2}-i\Delta)^{2}-\left(\beta\frac{\gamma}{2}e^{i\phi}\right)^{2}\right|^{2}},\nonumber\\
 & \equiv \left|T_{1}\right|^{2}I_{3}-2\im\sbkt{ T_{1}T_{2}^{*}} \beta\left(\frac{\gamma}{2}\right)I_{2}+\left|T_{1}-T_{2}\beta\right|^{2}\left(\frac{\gamma}{2}\right)^{2}I_{1}.
 \label{I0}
\end{align}

where we can simplify  the integrals $I_1$, $I_2$ and $I_3$ as follows:
\begin{align}
I_{1} & =\int_{0}^{\infty}d\omega\frac{1}{\left|(\frac{\gamma}{2}-i\Delta)^{2}-\left(\beta\frac{\gamma}{2}e^{i\phi}\right)^{2}\right|^{2}}=\frac{\pi}{2}\frac{1}{\left(\frac{\gamma}{2}\right)^{3}}\frac{1}{\left(1-\left(\beta\cos\phi\right)^{2}\right)\left(1+\left(\beta\sin\phi\right)^{2}\right)}\\
I_{2} & =\int_{0}^{\infty}d\omega\frac{\Delta}{\left|(\frac{\gamma}{2}-i\Delta)^{2}-\left(\beta\frac{\gamma}{2}e^{i\phi}\right)^{2}\right|^{2}}   =-\frac{\pi}{4}\frac{1}{\left(\frac{\gamma}{2}\right)^{2}}\frac{\beta^{2}\sin2\phi}{\left(1-\left(\beta\cos\phi\right)^{2}\right)\left(1+\left(\beta\sin\phi\right)^{2}\right)},\\
I_{3} & =\int_{0}^{\infty}d\omega\frac{\Delta^{2}}{\left|(\frac{\gamma}{2}-i\Delta)^{2}-\left(\beta\frac{\gamma}{2}e^{i\phi}\right)^{2}\right|^{2}}=\frac{\pi}{2}\frac{1}{\left(\frac{\gamma}{2}\right)}\frac{1-\beta^{2}\cos2\phi}{\left(1-\left(\beta\cos\phi\right)^{2}\right)\left(1+\left(\beta\sin\phi\right)^{2}\right)}.
\end{align}

Substituting the above  in Eq.\,\eqref{I0}, we get
\begin{equation}
I_{0}  =\frac{\pi}{\gamma}\frac{\left|T_{1}\right|^{2}\left[1-\beta^{2}\cos2\phi\right]+\im\sbkt{ T_{1}T_{2}^{*}} \beta^{3}\sin2\phi+\left|T_{1}-T_{2}\beta\right|^{2}}{\left(1-\left(\beta\cos\phi\right)^{2}\right)\left(1+\left(\beta\sin\phi\right)^{2}\right)}.
\end{equation}

Plugging this result back into Eq.\,\eqref{PL} and \eqref{PR} and considering $4\pi\left|g(\omega_{0})\right|^{2}=\beta\gamma$ yields
\begin{align}
P_{R}= & \frac{1}{2}\frac{\beta}{1+\left(\beta\sin\phi\right)^{2}}\left[\frac{\left(1-\beta+\left(\beta\sin\phi\right)^{2}\right)\left(1-\beta\cos\Delta\phi\cos\phi\sin2\theta\right)}{1-\left(\beta\cos\phi\right)^{2}}+\cos\left(\Delta\phi+\phi\right)\sin2\theta+2\beta\sin^{2}\theta\sin^{2}\phi\right],\\
P_{L}= & \frac{1}{2}\frac{\beta}{1+\left(\beta\sin\phi\right)^{2}}\left[\frac{\left(1-\beta+\left(\beta\sin\phi\right)^{2}\right)\left(1-\beta\cos\Delta\phi\cos\phi\sin2\theta\right)}{1-\left(\beta\cos\phi\right)^{2}}+\cos\left(\Delta\phi-\phi\right)\sin2\theta+2\beta\cos^{2}\theta\sin^{2}\phi\right].
\end{align}

which depends on the four parameters: $\beta$, $\theta$, $\phi$, and $\Delta\phi$. We use the above equations to obtain the total probability of emitting into the waveguide $P_\mr{tot}$  and the directionality parameter $\chi$ is given by  Eqs.\,\eqref{Eq:chi} and \eqref{eq:ptot}.

\subsection{Optimum directionality}
 We give the  parameter values that optimize the directionality parameter  given by Eq.\,\eqref{Eq:chi}. The value of $\phi$ that maximizes directionality is $\phi=\frac{2n+1}{2}\pi$ such that $\sin\phi=\pm1$. The  directional parameter for the optimal value of $\phi $ reads: \eqn{\chi=-\frac{\sin\Delta\phi\sin2\theta\pm\beta\cos2\theta}{1+\beta^{2}}.
\label{Eq:chiopt1}}

Considering that the value of $\beta $ is fixed for a given physical system, we find the global optimum over the two remaining parameters ($\theta$, and $\Delta\phi$), yields:
\begin{align*}
\beta\sin2\theta\mp\cos2\theta\sin\Delta\phi & =0\\
\mp\sin2\theta\cos\Delta\phi & =0,
\end{align*}
which gives the optimum values of of $\Delta \phi =\bkt{n+\frac{1}{2}}\pi$
and  $\theta=\pm\frac{1}{2}\arctan\frac{1}{\beta}$. This shows that the optimum value of $\theta$ in general depends on the value of $\beta$, and it tends to $\theta\rightarrow\bkt{\frac{2n+1}{2}}\pi$ as $\beta\rightarrow0$.

\subsection{Directional Fisher Information }

The directional and non-directional quantum Fisher information are defined as:

\eqn{ \mathcal{F}_{D}(\varphi)=&P_{L}(\varphi)\left(\frac{\partial \log P_{L}(\varphi)}{\partial\varphi}\right)^{2}+P_{R}(\varphi)\left(\frac{\partial \log P_{R}(\varphi)}{\partial\varphi}\right)^{2}+P_\mr{out}(\varphi)\left(\frac{\partial \log P_{out}(\varphi)}{\partial\varphi}\right)^{2}\\
\mathcal{F}_{ND}(\varphi)=&P_\mr{tot}(\varphi)\left(\frac{\partial \log P_\mr{tot}(\varphi)}{\partial\varphi}\right)^{2}+P_\mr{out}(\varphi)\left(\frac{\partial \log P_\mr{out}(\varphi)}{\partial\varphi}\right)^{2}
}

The difference between the two can be found as:

\begin{align}
\mathcal{F}_{D}(\varphi)-\mathcal{F}_{ND}(\varphi) & =P_{L}(\varphi)\left(\frac{\partial \log P_{L}(\varphi)}{\partial\varphi}\right)^{2}+P_{R}(\varphi)\left(\frac{\partial \log P_{R}(\varphi)}{\partial\varphi}\right)^{2}-P_{tot}(\varphi)\left(\frac{\partial \log P_{tot}(\varphi)}{\partial\varphi}\right)^{2}\\
& =P_{L}(\varphi)\left(\frac{\partial P_{L}(\varphi)}{\partial\varphi}\frac{1}{P_{L}(\varphi)}\right)^{2}+P_{R}(\varphi)\left(\frac{\partial P_{R}(\varphi)}{\partial\varphi}\frac{1}{P_{R}(\varphi)}\right)^{2}-P_{tot}(\varphi)\left(\frac{\partial P_{tot}(\varphi)}{\partial\varphi}\frac{1}{P_{tot}(\varphi)}\right)^{2}\\
& =\frac{1}{P_{L}(\varphi)+P_{R}(\varphi)}\left[\left(\frac{P_{R}(\varphi)}{P_{L}(\varphi)}\right)\left(\frac{\partial P_{L}(\varphi)}{\partial\varphi}\right)^{2}+\left(\frac{P_{L}(\varphi)}{P_{R}(\varphi)}\right)\left(\frac{\partial P_{R}(\varphi)}{\partial\varphi}\right)^{2}-2\frac{\partial P_{L}(\varphi)}{\partial\varphi}\frac{\partial P_{R}(\varphi)}{\partial\varphi}\right]\\
& =\frac{1}{P_{tot}(\varphi)}\left[\left(\sqrt{\frac{P_{R}(\varphi)}{P_{L}(\varphi)}}\right)\left(\frac{\partial P_{L}(\varphi)}{\partial\varphi}\right)-\sqrt{\frac{P_{L}(\varphi)}{P_{R}(\varphi)}}\left(\frac{\partial P_{R}(\varphi)}{\partial\varphi}\right)\right]^{2}.
\end{align}

Since all the probabilities are positive real numbers, this term is always positive, thus yielding:
\eqn{\mathcal{F}_{D}(\varphi)-\mathcal{F}_{ND}(\varphi)\geq0.} The equality is satisfied when $P_R  = P_L$, corresponding to  equal emission  probabilities in the left and right directions. 

\end{widetext}

\bibliographystyle{apsrev4-1}
\bibliography{wgqed}

\end{document}